\documentclass[journal=jacsat,manuscript=article]{achemso}

\usepackage[T1]{fontenc} 
\usepackage{textcomp}
\usepackage{upgreek}
\usepackage{array}

\usepackage{graphicx}
\usepackage{float}
\usepackage[none]{hyphenat}
\usepackage{amsmath}
\usepackage{epstopdf}
\usepackage{ragged2e}

\author{F. K. de Vries}
\altaffiliation{These authors contributed equally to this work.}
\author{J. Shen}
\altaffiliation{These authors contributed equally to this work.}
\email{j.shen-1@tudelft.nl}
\affiliation{QuTech and Kavli Institute of Nanoscience, Delft University of Technology, 2600 GA Delft, The Netherlands}

\author{ R.J. Skolasinski}
\affiliation{QuTech and Kavli Institute of Nanoscience, Delft University of Technology, 2600 GA Delft, The Netherlands}
\author{M. P. Nowak}
\affiliation{AGH University of Science and Technology, Academic Centre for Materials and Nanotechnology, al. A. Mickiewicza 30, 30-059 Krakow, Poland}
\author{D. Varjas}
\author{L. Wang}
\author{M. Wimmer}
\affiliation{QuTech and Kavli Institute of Nanoscience, Delft University of Technology, 2600 GA Delft, The Netherlands}

\author{J. Ridderbos}
\author{F. A. Zwanenburg}
\affiliation{NanoElectronics Group, MESA+ Institute for Nanotechnology, University of Twente, 7500 AE Enschede, The Netherlands}

\author{A. Li}
\altaffiliation{Current address: Beijing Key Lab of microstructure and Property of Advanced Materials, Beijing University of Technology, Pingleyuan No.100, 100024, Beijing, P. R. China}
\author{S. Koelling}
\affiliation{Department of Applied Physics, Eindhoven University of Technology, 5600 MB Eindhoven, the Netherlands}
\author{M. A. Verheijen}
\affiliation{Department of Applied Physics, Eindhoven University of Technology, 5600 MB Eindhoven, the Netherlands}
\affiliation{Philips Innovation Labs, 5656AE Eindhoven, the Netherlands}
\author{E. P. A. M. Bakkers}
\affiliation{Department of Applied Physics, Eindhoven University of Technology, 5600 MB Eindhoven, the Netherlands}

\author{L. P. Kouwenhoven}
\affiliation{QuTech and Kavli Institute of Nanoscience, Delft University of Technology, 2600 GA Delft, The Netherlands}
\alsoaffiliation{Microsoft Station Q Delft, 2600 GA Delft, The Netherlands}
\email{l.p.kouwenhoven@tudelft.nl}

\title{Spin-orbit interaction and induced superconductivity in an one-dimensional hole gas}

\keywords{spin-orbit interaction, mesoscopic superconductivity, nanowires, hole transport. \LaTeX}

\begin{document}

\newpage

\begin{abstract}
\noindent Low dimensional semiconducting structures with strong spin-orbit interaction (SOI) and induced superconductivity attracted much interest in the search for topological superconductors. Both the strong SOI and hard superconducting gap are directly related to the topological protection of the predicted Majorana bound states. Here we explore the one-dimensional hole gas in germanium silicon (Ge-Si) core-shell nanowires (NWs) as a new material candidate for creating a topological superconductor. Fitting multiple Andreev reflection measurements shows that the NW has two transport channels only, underlining its one-dimensionality. Furthermore, we find anisotropy of the Land\'{e} g-factor, that, combined with band structure calculations, provides us qualitative evidence for direct Rashba SOI and a strong orbital effect of the magnetic field. Finally, a hard superconducting gap is found in the tunneling regime, and the open regime, where we use the Kondo peak as a new tool to gauge the quality of the superconducting gap.
\end{abstract}
\vfill


\newpage


The large band offset and small dimensions of the Ge-Si core-shell NW leads to the formation of a high-quality one-dimensional hole gas \cite{Xiang_2006_nature,Conesa_2017}. Moreover, the direct coupling of the two lowest-energy hole bands mediated by the electric field is predicted to lead to a strong direct Rashba SOI \cite{Kloeffel_2011,Kloeffel_2017}. The bands are coupled through the electric dipole moments that stems from their wavefunction consisting of a mixture of angular momentum (L) states. On top of that the spin states of that wavefunction are mixed due to heavy and light hole mixing. Therefore an electric field couples via the dipole moment to the spin states of the system and causes the SOI. This is different from Rashba SOI which originates from the coupling of valence and conduction bands. The predicted strong SOI is interesting for controlling the spin in a quantum dot electrically \cite{NadjPerge_2010,Kloeffel_2013}. Combining this strong SOI with superconductivity is a promising route towards a topological superconductor \cite{Alicea_2012, Maier_2014}. Signatures of Majorana bound states (MBSs) have been found in multiple NW experiments \cite{Mourik_2012,Deng_2016}. An important intermediate result is the measurement of a hard superconducting gap \cite{Chang_2015,Gul_2017}, which ensures the semiconductor is well proximitized as is needed for obtaining MBSs.

Here we study a superconducting quantum dot in a Ge-Si NW. The scanning and transmission electron microscopy images of the device (Fig.\,1a and Fig.\,1b) show a Josephson junction of $\sim$\,170\,nm length. The quantum dot is formed in between the contacts. The NW has a Ge core with a radius of 3\,nm. The Ge crystal direction is found to be [110], in which hole mobilities up to 4600\,cm$^2$/ Vs are reported \cite{Conesa_2017}. The elemental analysis in Fig.\,1c reveals a pure Ge core with a 1\,nm Si shell and a 3\,nm amorphous silicon oxide shell around the wire. Superconductivity is induced in the Ge core by aluminium (Al) leads \cite{Xiang_2006} and, crucially, the device is annealed for a short time at a moderate temperature \cite{Su_2016,Ridderbos_2018}. We believe that the high temperature causes the Al to diffuse in the wire, therefore enhancing the coupling to the hole gas. Note that we do not diffuse the Al all the way through, since we pinch off the wire (Supplementary Fig.\,1) and there is no Al found in the elemental analysis (Fig.\,1c). Two terminal voltage bias measurements are performed on this device in a dilution refrigerator with an electron temperature of $\sim$\,50\,mK. 


\begin{figure*}[!t]
	\includegraphics[width=6in]{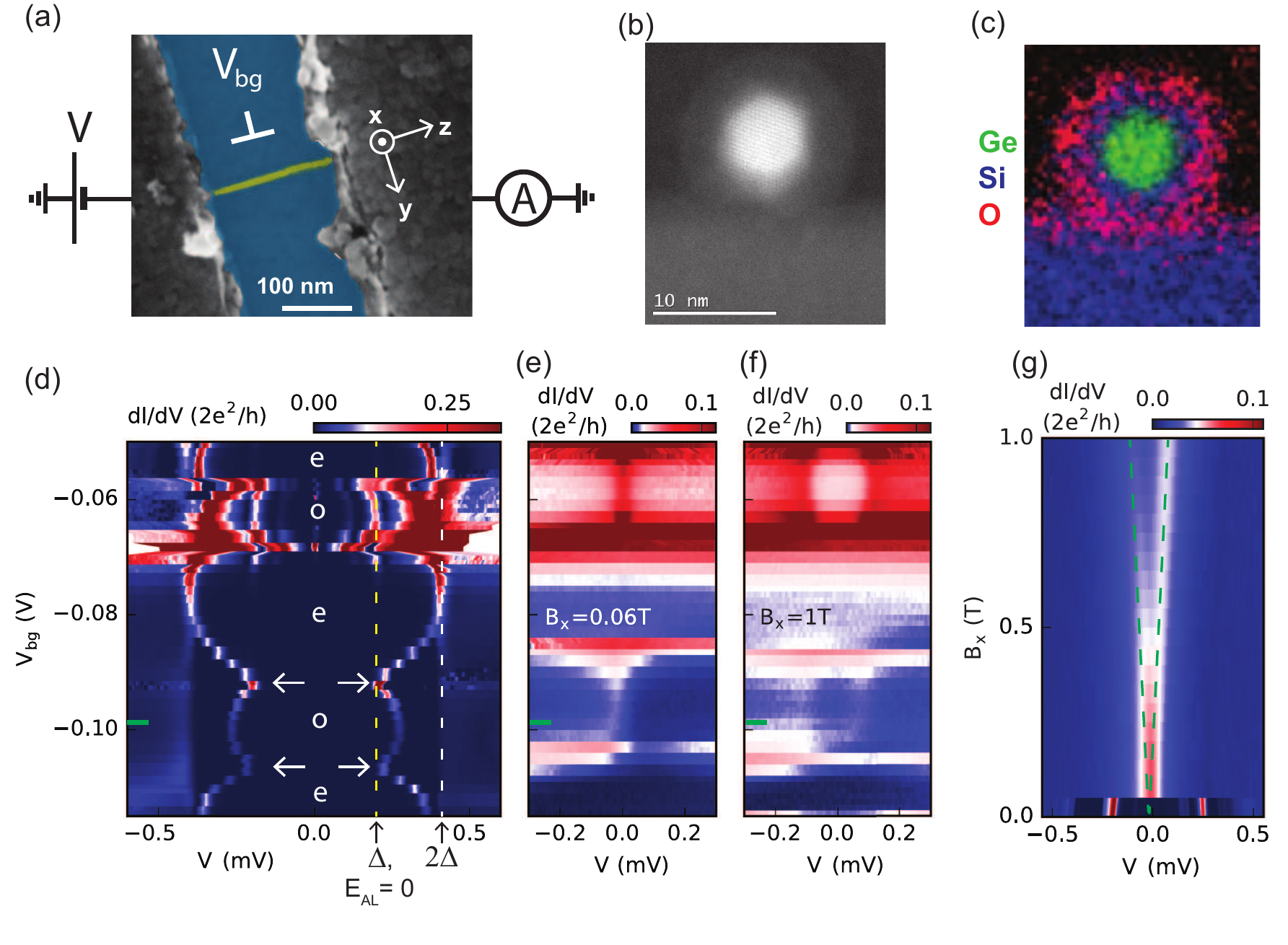}
	\justify
	\caption{(a) False colored scanning electron microscope image of the device with the NW (yellow) with aluminium contacts (grey) on a Si/SiN$_x$ wafer (blue). The magnetic field axes, voltage bias measurement setup and global bottom gate are indicated. (b) Transmission electron microscope (TEM) image of the cross section of the NW. (c) Energy dispersive X-ray spectroscopy of the area displayed in (b). The colors represent different elements, Ge is green, Si blue and oxygen (O) red, respectively. The Ge-Si core-shell wire is capped by a SiO$_x$ shell. (d) Voltage bias tunneling spectroscopy measurement of the superconducting quantum dot as the bottom gate voltage $V_{bg}$ is altered. The superconducting gap, an Andreev level (AL) and multiple Andreev reflections appear as peaks in differential conductance ($dI/dV$). $\Delta$ and $2\Delta$ are marked by the dashed yellow and white line, respectively. The even or odd occupation is indicated and the kink in the observed Andreev level is highlighted by the arrows. (e-f) Same measurement as (d) with a magnetic field $B$ applied perpendicular to the substrate ($x$-direction) of 60\,mT and 1\, T, respectively. A zero bias Kondo peak is observed as the quantum dot is occupied by an odd number of electrons. At $B$\,=\,1\,T, the resonance is split due to the Zeeman effect. (g) Linear splitting of the Kondo peak at $V_{bg}$\,=\,-0.098\,V as a function of $B$. The Zeeman effect splits the spinful Kondo peak, which is indicated by the dashed green line. 
	}
	\label{GeSi:fig1}
\end{figure*}

To perform tunneling spectroscopy measurements the bottom gate voltage $V_{bg}$ is used to vary the barriers of the quantum dot and alter the density of the holes as well. From a large source-drain voltage $V$ measurement (Supplementary Fig.\,1), we estimate a charging energy $U$ of 12\,meV, barriers' asymmetry of $\Gamma_1/\Gamma_2$\,=\,0.2-0.5, where $\Gamma_{1 (2)}$ is the coupling to the left (right) lead, and a lever arm of 0.3\,eV/V. In Fig.\,1d, the differential conductance $dI/dV$ as a function of a $V$ versus $V_{bg}$ reveals a superconducting gap (2$\Delta$\,=\,380\,$\mu$eV) and several Andreev processes within this window. Additionally, an even-odd structure shows up  in both the superconducting state at low $V$ and normal state at high $V$ , which is related to the even or odd parity of the holes in the quantum dot. The even-odd structure persists as we suppress the superconductivity in the device by applying a small magnetic field (60\,mT) perpendicular to the substrate (Fig.\,1e). A zero bias peak appears when the quantum dot has odd parity. This is a signature of the Kondo effect \cite{Goldhaber_1998,Cronenwett_1998}. When increasing the magnetic field to 1\,T, the Kondo peak splits due to the Zeeman effect by $2g\mu_BB$. The energy splitting of the two levels is linear as shown in Fig.\,1g, and thus can be used to extract a Land\'{e} g-factor $g$ of 1.9. In the remainder of the letter we will discuss the three magnetic field regimes of Fig.\,1d-f (0\,T, 60\,mT and\,1T, respectively) in more detail.

\begin{figure*}[!t]
	\centering
	\includegraphics[width=5in]{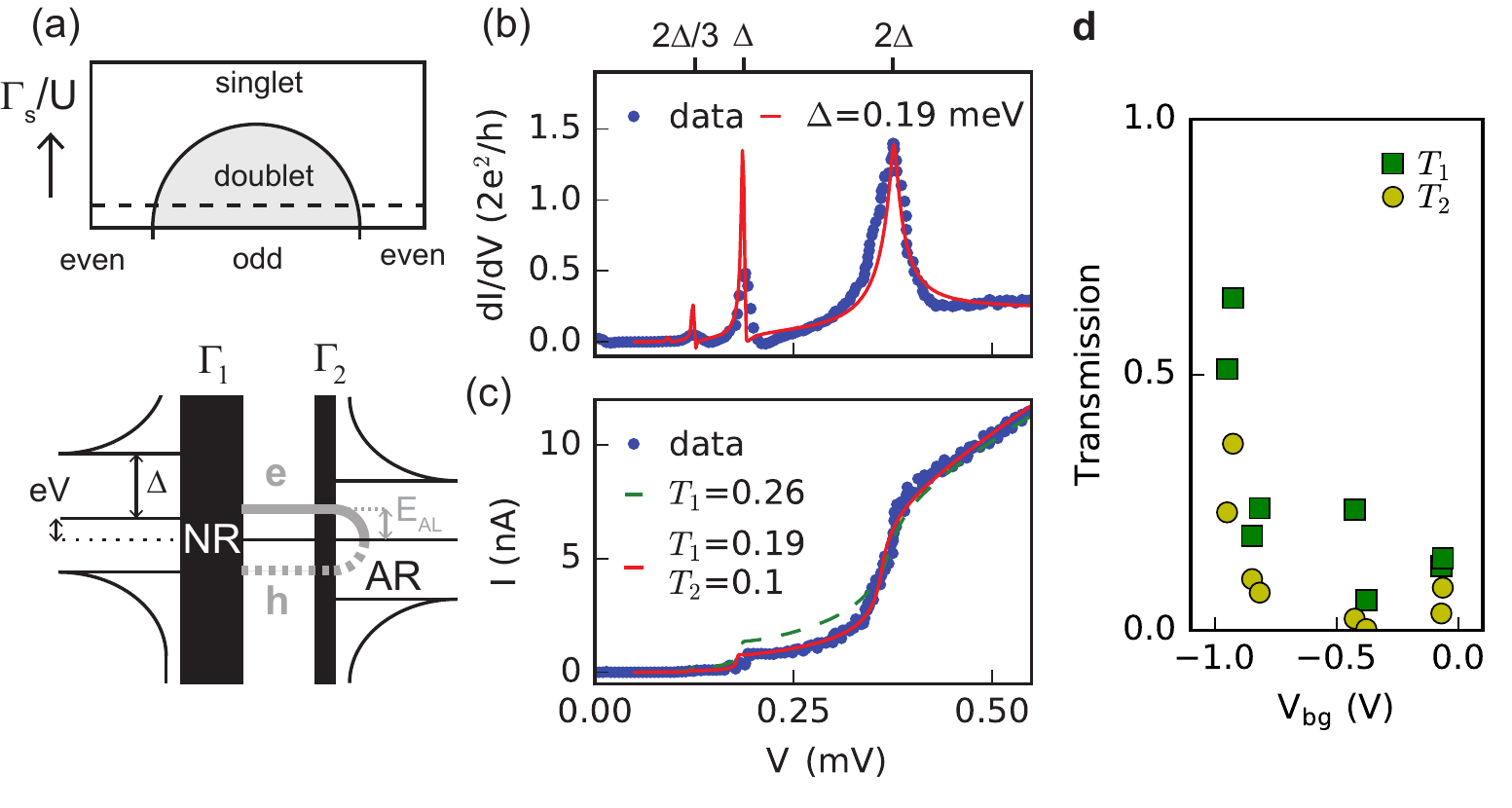}
	\justify
	\caption{(a) In the top panel a phase diagram of the ground state in the superconducting quantum dot is sketched. Because of the large charging energy $U$ compared to the coupling to the superconducting reservoir $\Gamma_s$, we expect to trace the dashed line. The bottom panel shows the Andreev level (dashed grey line) with energy $E_{AL}$ that is formed by Andreev Reflection (AR) at one side and Normal Reflection (NR) at the other side of the dot. The reflection processes are different due to asymmetric barriers $\Gamma_1$ and $\Gamma_2$, indicated as the barrier width. The density of states in the NW is probed by the superconductor on the left side by doing voltage bias tunneling spectroscopy. (b) Tunneling spectroscopy measurement at $V_{bg}$\,=\,-0.85\,V. The first and second order multiple Andreev reflection are observed. A two mode model fits the data well with $\Delta$\,=\,190\,$\mu$eV. (c) Measured current of (b). The data is fitted with a single and two mode model. The latter resembles the data better and is therefore used to extract transmission values. (d) Transmission of the first and second mode, $T_1$ and $T_2$, extracted from the fit of multiple Andreev reflections different $V_{bg}$. The transmission increases significantly below $V_{bg}$\,=\,-0.8\,V.   
	}
	\label{GeSi:fig2}
\end{figure*}

The resonance that disperses with $V_{bg}$ in Fig.\,1d is an Andreev Level (AL), which is the energy transition from the ground to the excited state in the dot \cite{Deacon_2010,Lee_2014}. The ground state of the dot switches between singlet and doublet if the occupation in the dot changes, as sketched in the phase diagram in the top panel of Fig.\,2a. Since our charging energy is large, we trace the dashed line in the phase diagram. The AL undergoes Andreev reflection at the side of the quantum dot with large coupling ($\Gamma_2$) and normal reflection at the opposite side that has lower coupling ($\Gamma_1$), as schematically drawn in bottom panel of Fig.\,2a. The superconducting lead with the low coupling serves as a tunneling spectroscopy probe of the density of states. To be more precise, the coherence peak of the superconducting gap is the probing the Andreev level energy $E_{AL}$. For example if $E_{AL}=0$ we measure it at $eV=\Delta$, the resonance thus has an offset of $\pm\Delta$ in the measurement in Fig\,1d. The ground state transition is visible as a kink of the resonance at $V$\,=\,$\Delta$ at $V_{bg}$\,=\,-0.09\,mV and -0.11\,mV. At more negative $V_{bg}$ the coupling of the hole gas to the superconducting reservoirs is strongly enhanced. 
This eventually leads to the observation of both the DC and AC Josephson effects (Supplementary Fig.\,2).

\begin{figure*}[!t]
	\includegraphics[width=6.8in]{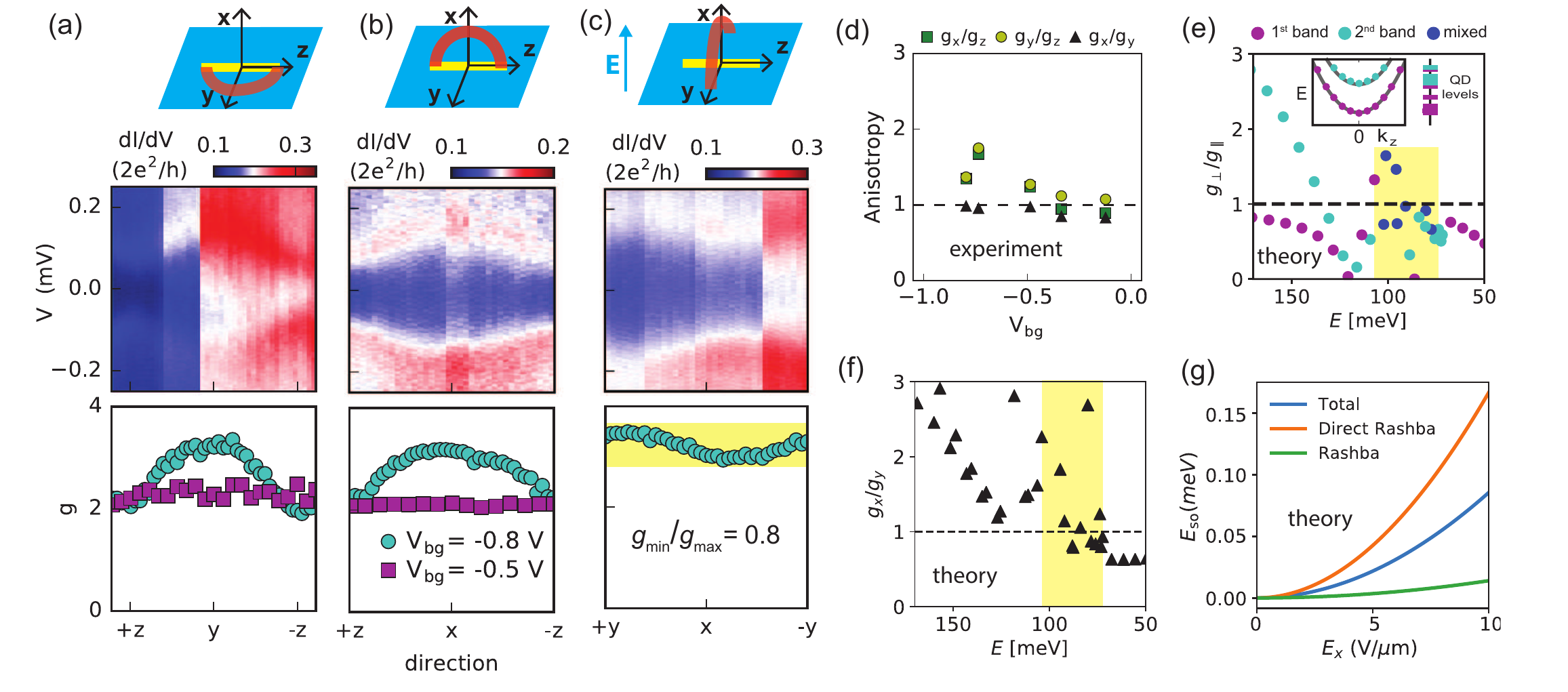}
	\justify
	\caption{(a-c) Rotations of the magnetic field with a 0.9\,T magnitude in the $yz$-, $xz$- and $xy$-plane, respectively, at $V_{bg}$\,=\,-0.79\,V. The upper panel shows the schematic of the device and the magnetic field rotation performed. The differential conductance data is plotted in the center panel, the splitting of the Kondo peak changes as the angles are swept. The sudden changes in conductance are due to small switches in $V_{bg}$. The lower panel shows the extracted $g$ of the center panel in cyan and $g$ at $V_{bg}$\,=\,-0.5 V in magenta. For the $xy$-plane the anisotropy is highlighted and calculated. (d) Summary of the measured anisotropies of $g$ at different $V_{bg}$. (e) Simulation result of the quantum dot. The anisotropy of $g_\parallel$ and $g_\bot$ changes as the Fermi energy is altered. The colors represent the band where the quantum dot level predominantly stems from. The highlighted part shows similar behaviour in the anisotropy values as the data in (d). The inset depicts a schematic representation of the energy ordering of the quantum dot levels originating from two bands along the NW. (f) Simulation as (e), now with an applied electric field of 10\,V/$\mu$m. The SOI causes anisotropy with respect to the electric field direction as $g_x$ is pointed perpendicular and $g_y$ parallel to the electric field. The anisotropy increases as the Fermi level is raised. The same range as in (e) is highlighted. (g) Simulated spin-orbit energies in the quantum dot as a function of electric field along the $x$-direction. The direct Rashba term is the leading contribution.
}
	\label{GeSi:fig3}
\end{figure*}

In the upper part of Fig.\,1d we measure multiple Andreev reflection (MAR): resonances at integer fractions of the superconducting gap. Fig.\,2b presents a line trace at $V_{bg}$\,=\,-0.85\,V that shows the gap edge and first- and second- order Andreev reflection. Fitting the differential conductance \cite{Averin_1995,Kjaergaard_2017} (see Supplementary) allows us to extract $\Delta$=\,190\,$\mu$eV, close to the bulk gap of Al. We also fit the measured current to extract the transmission of the spin degenerate longitudinal modes in the NW (Fig.\,2c) \cite{Scheer_1997,Goffman_2017}. The two-mode fit resembles the data better than the single mode fit. Therefore the first provides us an estimate for the transmission in the two modes, $T_1$ and $T_2$. We interpret the two modes as two semiconducting bands in the NW. The MAR fitting analysis is repeated at different $V_{bg}$ and the resulting $T_1$ and $T_2$ are plotted in Fig.\,2d. The strong increase of the transmission below $V_{bg}$\,=\,-0.8\,V is attributed to the increase of the Fermi level, and $\Gamma_1$ and $\Gamma_2$.

The Land\'{e} g-factor $g$ is investigated further by measuring the Kondo peak splitting as a 0.9\,T magnetic field is rotated from $y$- to $z$-, $x$- to $z$- and $x$- to $y$-direction as presented in the second row of Fig.\,3a-c. Interestingly, we find a strong anisotropy of the Kondo peak splitting and accordingly of $g$ at $V_{bg}$\,=\,-0.79\,V  (bottom row Fig.\,3a-c). Both directions perpendicular to the NW show a strongly enhanced $g$.  Similar anisotropy has been reported before in a closed quantum dot, where $g$ is even quenched in the $z$-direction \cite{Maier_2013,Brauns_2016_efield,Brauns_2016_pauli}. In our experiment the highest $g$ of 3.5 is found when the magnetic field is pointed perpendicular to the NW, and almost perpendicular to the substrate.

On the contrary, at a $V_{bg}$\,=\,-0.5\,V we find an isotropic $g$ (bottom row of Fig.\,3a-c), all of which have a value of around 2. The anisotropies at different $V_{bg}$ are summarized in Fig.\,3d. The strong anisotropy seems to set in around $V_{bg}$\,=\,-0.7\,V. This sudden transition from isotropic to anisotropic $g$, which has not been observed before in a quantum dot system, is correlated with the increase in transmission in Fig.\,2d. We speculate that the change from isotropic to anisotropic behaviour is related to the occupation of two bands in the NW. To test this hypothesis and get an understanding of the origin of the anisotropy we theoretically model the band structure of our NW and focus on the two lowest bands. 

We use the model described in Ref. \citenum{Kloeffel_2017} and apply it to our experimental geometry (see Supplementary for details). Simulating the device as an infinite wire we first consider the anistropy of $g$ between the directions parallel and perpendicular to the NW. We find that there are two contributions to the anisotropy: the Zeeman and the orbital effect of the magnetic field \cite{Nijholt_2016,Winkler_2017}. The anisotropy of the Zeeman component is similar for the two lowest bands, where for the orbital part the anisotropy differs strongly. The anisotropy of the total $g$ therefore shows a strong difference for the two lowest bands (Supplementary Figs. 5-6). This agrees qualitatively with earlier predictions\cite{Kloeffel_2011}, but we find additionally that strain lifts the quenching of $g$ along the NW such that $g_\parallel/g_\bot \sim 2$, in agreement with our measurements. From these observations we conclude that the observed isotropic and anisotropic $g$ with respect to the NW-axis is due to the orbital effect. 

In addition, we include the confinement along the NW, such that a quantum dot is formed and the energy levels are quantized in the $z$-direction. Besides the lowest energy states studied before,\cite{Maier_2013,Kloeffel_2013} we also consider a large range of higher quantum dot levels. In the regime where two bands are occupied we observe that the quantum dot levels originating from the first and second band have a unique ordering as a function of Fermi energy, this situation is sketched in the inset of Fig.\,3e. We also find that some of the quantum dot levels are a mixture of the two bands (Supplementary Fig.\,8), resulting in a different anisotropy for each quantum dot level. In the simulation results (Fig.\,3e and Supplementary Fig.\,9) the anisotropy values are colored according to the band they predominantly originate from. To compare the simulation with the measured data we note that a more negative $V_{bg}$ in the experiment increases the Fermi level for holes $E$. In the simulation we observe a regime in $E$ (highlighted in Fig.\,3e), where the anisotropy $g_\bot$/ $g_\parallel$ is around 1 and goes up towards 2 as $E$ increases. This behaviour qualitatively resembles the measurement of $g_x/g_z$ and $g_y/g_z$ in Fig.\,3d.

Now we turn to the magnetic field rotation in the $xy$-plane, the two directions perpendicular to the NW that are parallel and perpendicular to the electric field induced by the bottom gate. The measured anisotropy is $g_{min}$/$g_{max}$\,=\,0.8 (Fig.\,3c). The maximum $g$ of 3.5 is just offset of the $y$-direction, which is almost parallel to the electric field. This anisotropy with respect to the electric field direction is a signature of SOI \cite{Maier_2013,Brauns_2016_efield}. As discussed before, the Ge-Si NWs are predicted to have both Rashba SOI and direct Rashba SOI \cite{Kloeffel_2011,Kloeffel_2013}. The electric field could also cause anisotropy via the orbital effect or geometry, due to an anisotropic wavefunction. However we can rule that out since our simulations show that the wavefunction does not significantly change as electric field is applied (Supplementary Fig.\,7). In the simulation (Fig.\,3f) with a constant electric field of 10\,mV/$\mu$m, we observe anisotropy of $g$ parallel ($g_x$) and perpendicular($g_y$) to the electric field. Similar to our data the anisotropy starts below 1 and goes to 1 as the Fermi level is increased. The spread in the anisotropy values is due to the mixing of the bands for each quantum dot level. Furthermore we calculated the magnitude of the Rashba and direct contribution to the SOI and find the direct Rashba SOI is dominating in the small diameter nanowires of our study (Fig.\,3g). This agrees with the effective Hamiltonian derived in Ref.~\citenum{Kloeffel_2011}, which predicts that the direct Rashba SOI dominates in NWs with a Ge core of 3\,nm radius. To summarize, we observe anisotropy with respect to the electric field direction that is caused by SOI, which is likely for the largest part due to the direct Rashba SOI.

\begin{figure}[!t]
	\centering
	\includegraphics[width=5in]{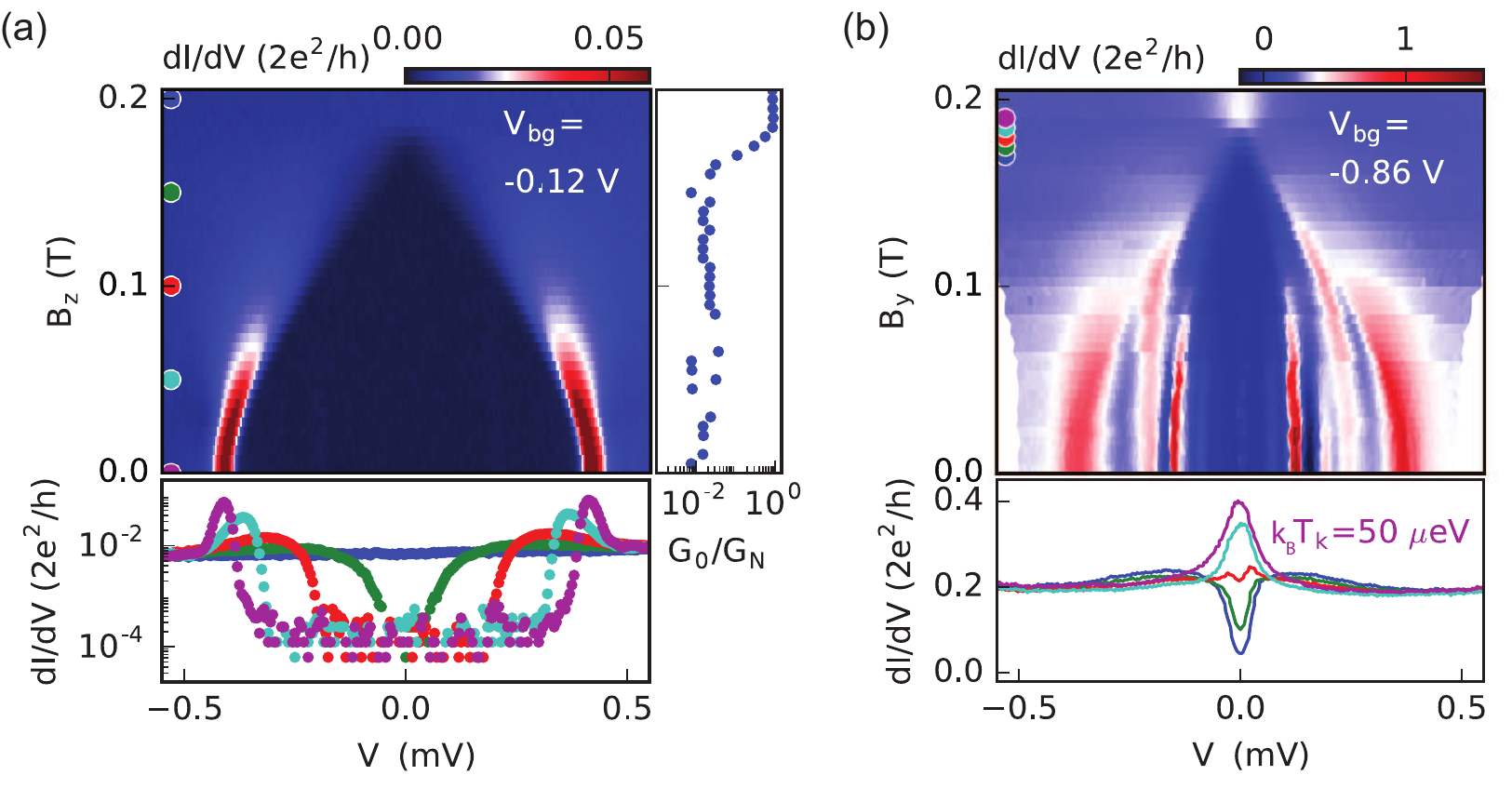}
	\justify
	\caption{(a) The superconducting gap closes as $B$ is ramped up in the $z$-direction. The line traces below are taken at 50\,mT intervals and show the induced superconducting gap. The vertical line trace shows the conductance at $V$\,=\,0\,V normalized to the conductance extracted at $V$\,=\,0.5\,mV. A two orders of magnitude conductance suppression is observed. (b) The superconducting gap closes and a Kondo peak appears as the magnetic field is increased in the $y$-direction. The resonances within the gap stem from Andreev processes. The line traces depict the transition from the superconducting gap to the Kondo peak, which takes place from 170\,mT to 190\,mT (5\,mT step). From the pink trace a Kondo energy $k_BT_K$ of 50\,$\mu$eV is extracted with an Lorentzian fit.
	}
	\label{GeSi:fig4}
\end{figure}

Finally, in Fig.\,4 we take a detailed look at the superconducting gap as a function of magnetic field. We find the critical magnetic field $B_c$ for different directions:  $B_{c,z}$\,=\,220\,mT (Fig.\,4a), $B_{c,y}$\,=\,220\,mT (Fig.\,4b), and $B_{c,x}$\,=\,45\,mT(Fig.\,1g and Supplementary\,Fig.\,3), consistent with an Al thin film. In the tunneling regime at $V_{bg}$\,=\,-0.12\,V, we observe a clean gap closing (Fig.\,4a). The conductance inside the gap is suppressed by two orders of magnitude, signaling a low quasiparticle density of states in the superconducting gap. This large conductance suppression remains as the gap size decreases towards $B_c$ (bottom panel in Fig.\,4a). In the low conductance regime we thus measure a hard superconducting gap persisting up to $B_c$ in Ge-Si NWs. 

The closing of the superconducting gap in a higher conductance regime is presented in Fig.\,4b. Since the transmission is increased, Andreev reflection processes cause a significant conductance within the superconducting gap \cite{BTK_1982}. Therefore the conductance suppression in the gap becomes an ill-defined measure of the quasiparticle density of states and with that the quality of the induced superconductivity. However, here we can use the Kondo peak to examine the quasiparticle density of states in the superconducting gap. The Kondo peak is formed by coupling through quasiparticle states within the window of the Kondo energy ($k_BT_K$). In the regime where $k_BT_K\,\leq\,\Delta$, the existence and size of the Kondo peak is then an indication of the quasiparticle density of states inside the superconducting gap \cite{Buitelaar_2002,Lee_2012}. In our measurement $\Delta$ is indeed than $k_BT_K$ up to a magnetic field $B$\,=\,170\,mT (see the blue and magenta line traces in bottom panel of Fig.\,4b). Since in the measurement the Kondo peak only arises once the gap is fully closed, we have a low quasiparticle density of states within the superconducting gap. This supports our observation of a hard superconducting gap up to $B_c$. It also illustrates a new way of gauging whether the superconducting gap is hard in a high conductance regime.

Combining all three magnetic field regimes of Fig. 2-4, we observed: Andreev levels showing a ground state transition; SOI from the coexistence of two bands in Ge-Si core-shell NWs; and a hard superconducting gap. The combination and correlation of these observations  is a crucial step for exploring this material system as a candidate for creating a one-dimensional topological superconductor.\\

\subsection{Associated Content}
\textbf{Supporting Information}\\
The supporting information entails extra experimental data and a description and intermediate results of the band structure calculations. 

\subsection{Author Information}
\textbf{Author contributions}\\
F.K.d.V. and J.S. designed the experiment, fabricated the devices and performed the measurements. M.P.N. and M.W. did the MAR fitting. R.S., D.V., L.W., and M.W. performed band structure calculations. J.R. and F.Z. contributed to the discussions of data. A.L. and E.P.A.M.B. grew the material. S.K., and M.A.V. did TEM analysis. L.P.K. and J.S. supervised the project. F.K.d.V., J.S. and R.S. wrote the manuscript. All authors commented on the manuscript.\\
\textbf{Notes}\\
The authors declare no competing financial interests.

\subsection{Acknowledgements}
The authors thank M. C. Cassidy for fruitful discussions about the fabrication and Y. Ren for help with the growth. This work has been supported by funding from the Netherlands Organisation for Scientific Research (NWO), Microsoft Corporation Station Q and the European Research Council (ERC HELENA 617256 and ERC Starting Grant 638760). We acknowledge Solliance, a solar energy R\&D initiative of ECN, TNO, Holst, TU/e, imec and Forschungszentrum J\"ulich, and the Dutch province of Noord-Brabant for funding the TEM facility. M.P.N. acknowledges support by the National Science Centre, Poland (NCN).

\providecommand{\latin}[1]{#1}
\makeatletter
\providecommand{\doi}
  {\begingroup\let\do\@makeother\dospecials
  \catcode`\{=1 \catcode`\}=2 \doi@aux}
\providecommand{\doi@aux}[1]{\endgroup\texttt{#1}}
\makeatother
\providecommand*\mcitethebibliography{\thebibliography}
\csname @ifundefined\endcsname{endmcitethebibliography}
  {\let\endmcitethebibliography\endthebibliography}{}

\setcounter{figure}{0}

\makeatletter 
\renewcommand{\thefigure}{\normalsize\textbf{S\arabic{figure}}}
\renewcommand{\thetable}{\normalsize \textbf{S\arabic{table}}}
\makeatother

\renewcommand\figurename{\normalsize \textbf{Figure}}
\renewcommand\tablename{\normalsize\textbf{Table}}

\newpage

\begin{center}
\LARGE{ \huge Supplementary Material \\[5mm] \LARGE Spin-orbit interaction and induced superconductivity in an one-dimensional hole gas}\\[1cm]
\linespread{1.5}
\large \noindent F. K. de Vries, J. Shen, R.J. Skolasinski, M. P. Nowak,  D. Varjas, L. Wang, M. Wimmer, J. Ridderbos, F. A. Zwanenburg, A. Li, S. Koelling, M. A. Verheijen, E. P. A. M. Bakkers and L. P. Kouwenhoven\\[2cm]
\end{center}

\normalsize
\clearpage

\begin{figure}[!ht]
	\centering
	\includegraphics[width=.5\columnwidth]{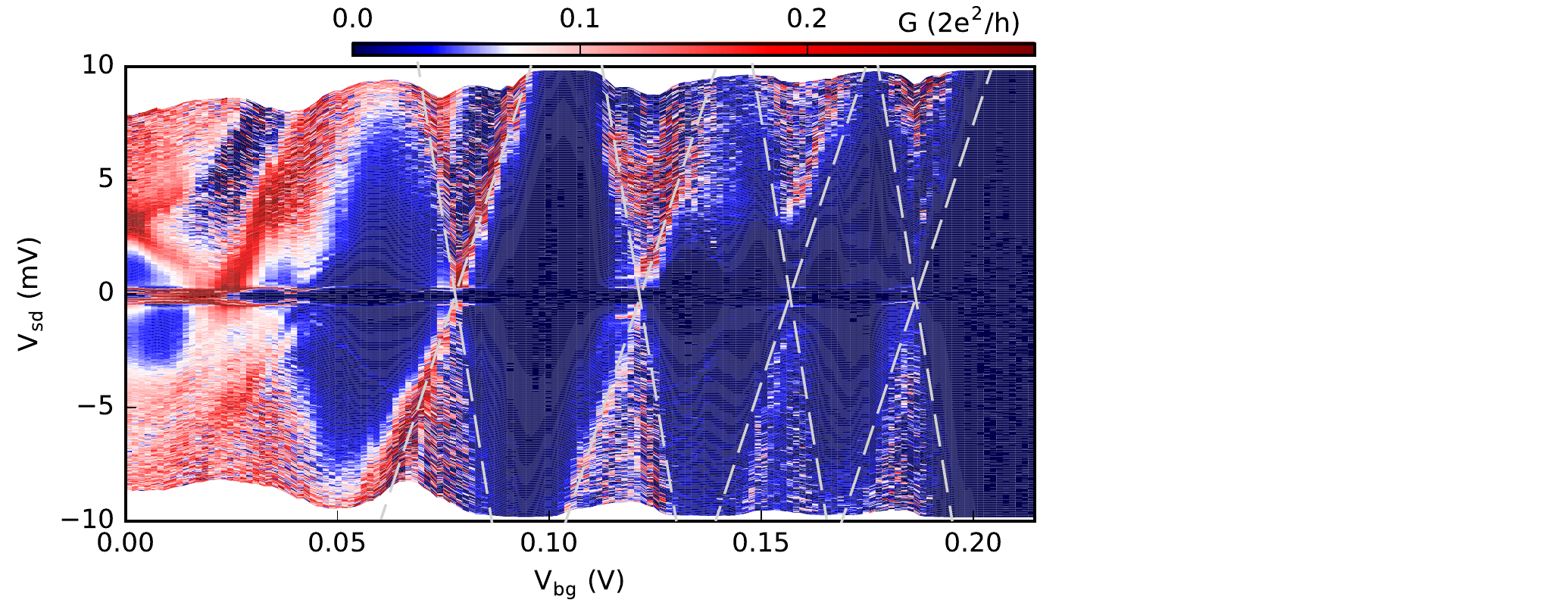}
	\caption{Large $V_{sd}$ tunneling spectroscopy measurement of the superconducting quantum dot. The differential conductance ($dI/dV$) as a function of $V_{bg}$ reveals Coulomb diamonds that are highlighted by the dashed lines. From $V_{bg}$\,=\,0.2\,V on the hole transport is pinched off. The charging energy of 12\,meV , barriers’ asymmetry of $\Gamma_1$/$\Gamma_2$\,=\,0.2-0.5 and lever arm of 0.3\,eV/V are estimated from this graph.
	}
	\label{fig:S1}
\end{figure}

\begin{figure}[H]
	\includegraphics[width=.8\columnwidth]{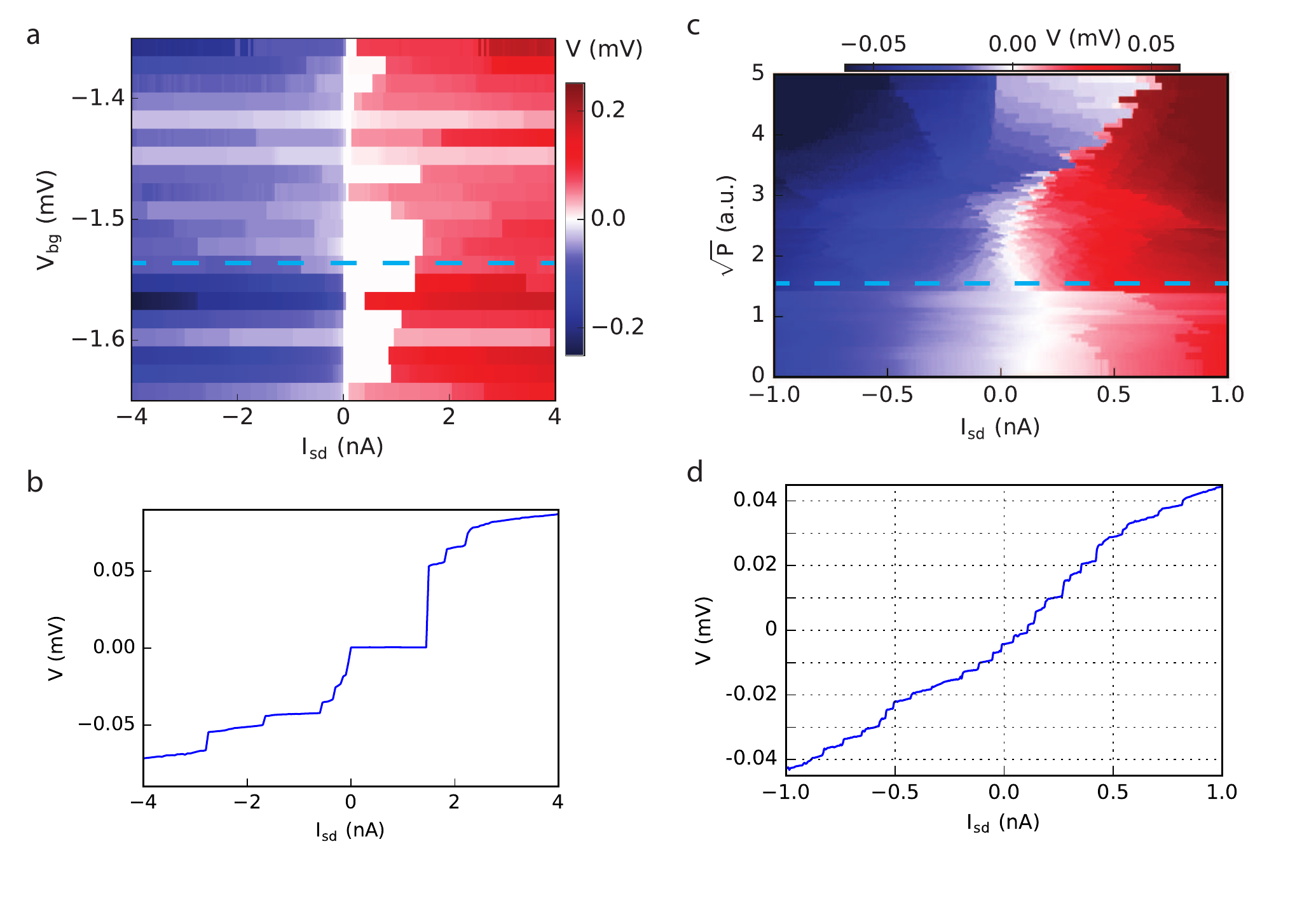}
	\caption{(a) Voltage measurement as a function of current bias $I_{sd}$ and $V_{bg}$. The measured voltage shows supercurrent as zero voltage plateaus. (b) Linetrace at $V_{bg}$\,=\,-1.3\,V. A switching current of 1.7\,nA is observed. (c) Voltage measurement as a function of current bias at $V_{bg}$\,=\, -1.53\,V while a microwave excitation is applied with a varying power at a frequency $f$ of 1.23 GHz. (d) A linetrace at $\sqrt{P}$\,=\,1.46 reveals Shapiro steps with stepsize corresponding to the frequency $V=f h/2e=2.5 \mu V$.
	}
	\label{fig:S2}
\end{figure}

\newpage
\begin{figure}[!ht]
	\includegraphics[width=.4\columnwidth]{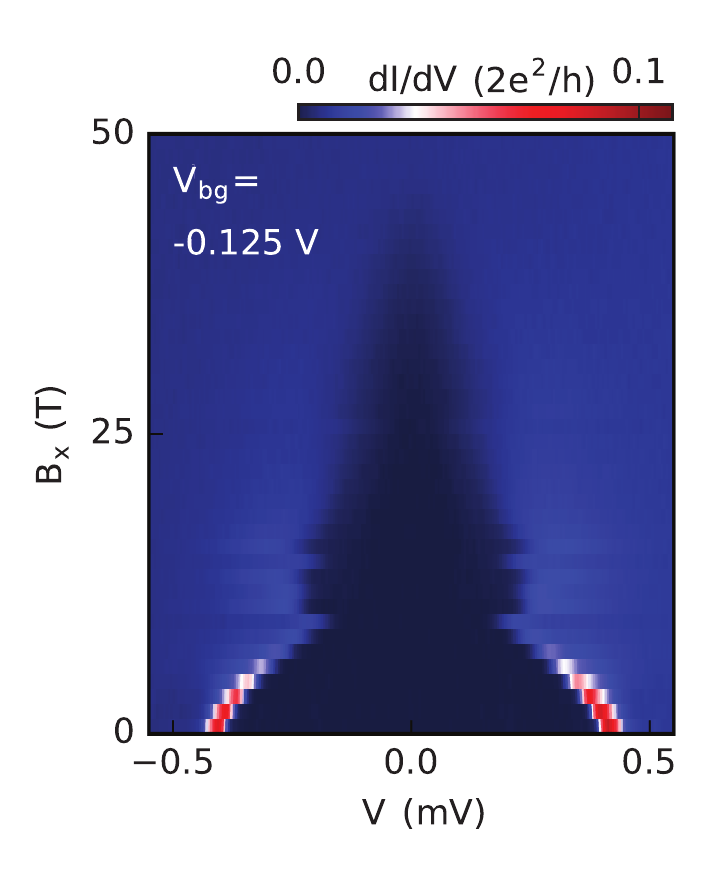}
	\caption{Tunneling spectroscopy measurement of the superconducting gap as a function of $B_x$, a magnetic field applied perpendicular to the substrate. The superconducting gap closes at a critical magnetic field of 45\,mT. At 10\,mT the gap closing seems to set in early. This dual gap closing is seen also in a similar experiment, where it is attributed to a superconducting material that is created during the annealing proces and that is composed of Al and Si \cite{Ridderbos_2018_SM}. 
	}
	\label{fig:S3}
\end{figure}
\newpage

\section{Calculation of multiple Andreev reflection and the fitting procedure}
The conductance and the current response of the voltage biased nanowire Josephson junction are calculated following the scattering approach introduced by Averin and Bardas in Ref. \cite{averin_ac_1995}. The model accounts for sequential Andreev reflections of electrons and holes accelerated by the voltage bias $V_{\mathrm{sd}}$ that propagate through the normal part of a SNS junction. The total DC current $I_{\mathrm{th}}(V_{\mathrm{sd}})$ of a multimode junction is obtained \cite{bardas_electron_1997} as a sum of the currents $I_{i}$ carried by individual modes of the transverse quantization
\begin{equation}
I_{\mathrm{th}}(V_{\mathrm{sd}}) = \sum_{i}^{N}I_{i}(V_{\mathrm{sd}}, T_i, \Delta),
\end{equation} 
where $T_i$ is the transmission probability of the $i$'th mode and $\Delta$ is the superconducting gap.

The transmission probabilities $T_i$ and the superconducting gap $\Delta$ of the measured nanowire junction are inferred by fitting the numerically obtained current to the experimental one through minimization of $\chi = \int[I_\mathrm{{exp}}(V_{\mathrm{sd}})-I_{\mathrm{th}}(V_{\mathrm{sd}})]^2 dV_{\mathrm{sd}}$. $N$ is a free parameter of the fitting procedure. We have checked that increase of $N$ above 2 results in the transmission probabilities $T_{i>2}=0$ evidencing the presence of only two conducting modes in the structure as described in the main text. The analogous procedure is performed for the differential conductance traces, obtained in the numerics by differentiation of the calculated current over the bias voltage.
\newpage

\section{Numerical calculations}

\subsection{Discussion of previous results and overview}

Our experimental data shows a g-factor with a gate-tunable anisotropy. A g-factor anisotropy for Ge-Si core-shell nanowires with a circular crosssection and in the absence of electric fields was predicted for the lowest two subbands in Ref.~\citenum{Kloeffel_Strongspinorbitinteraction_2011}: At $k_z=0$, the g-factors for the lowest subband were computed to be $g_z=0.12$ and $g_{x,y}=5.78$, for the second subband $g_z=3.12$ and $g_{x,y}=5.1$. Comparing to the experimentally measured values, we observe that (i) the computed anisotropy in the lowest subband is larger than in the second subband, whereas we observe a quenched anisotropy for lower densities, and (ii) the experimentally measured $g_z$ never drops below 2. A later numerical simulation including strain found for the lowest subband $g_z=1.8$ and $g_{x,y}=8.3$ \cite{Kloeffel_DirectRashbaspinorbit_2017}, i.e.~bringing the g-factor for field parallel to the wire closer to our experimental results. However, no results were given for the second subband there. Ref.~\citenum{Maier_2013_SM} discussed the electric field dependence of the g-factor anisotropy for the ground state in a quantum dot in the Ge-Si core-shell nanowire, and found that the anisotropy was quenched with increasing electric field due to the direct Rashba SOI. Again, this is opposite to our experimental observation that the anisotropy is quenched for small (absolute) gate voltages.

We can thus not directly interpret our results in terms of existing theory. For this reason we apply the model described in Ref.~\citenum{Kloeffel_DirectRashbaspinorbit_2017} to our experimental geometry and strain values. As we show below, strain can change g-factor values up to an order of magnitude and even reverse anisotropies. We also find that we need to consider excited quantum dot states to find agreement with the experimental data.

\subsection{Model for nanowire along [110]}
We use the Luttinger-Kohn Hamiltonian for holes that has been established for modelling Ge-Si core shell nanowires \cite{Kloeffel_Strongspinorbitinteraction_2011, Kloeffel_DirectRashbaspinorbit_2017}. Below we give this Hamiltonian in detail, the description was adapted from Ref.~\citenum{Kloeffel_DirectRashbaspinorbit_2017}. The bulk Hamiltonian of the Ge core is

\begin{equation}
	\label{eq-supp-hamiltonian}
	H = H_\textrm{LK} + H_\textrm{Z} + H_\textrm{dir} + H_\textrm{R} + H_\textrm{BP}\,,
\end{equation}
where $H_\textrm{LK}$ is the Luttinger-Kohn (LK) Hamiltonian, $H_\textrm{dir}$ the coupling to the electric field that is known to give rise to the direct Rashba spin-orbit interaction (SOI) \cite{Kloeffel_Strongspinorbitinteraction_2011}, $H_\textrm{R}$ is the indirect Rashba SOI due to coupling to other bands, and $ H_\textrm{BP}$ is the Bir-Pikus Hamiltonian describing the effects of strain.
The magnetic field is included through the Zeeman term $H_\textrm{Z}$, and the orbital effect. We consider the orbital effect of the field through kinetic momentum substitution.
We include a global ``$-$'' sign in our Hamiltonian such that hole states have a positive effective mass.
In the following we take a detailed look at each term in the Hamiltonian separately.

\paragraph{Luttinger-Kohn Hamiltonian}
We use the Luttinger-Kohn (LK) Hamiltonian~\cite{Luttinger_MotionElectronsHoles_1955, Luttinger_QuantumTheoryCyclotron_1956}

\begin{equation}
	\label{eq-supp-lk}
	H_{\textrm{LK}} = \frac{\hbar^2}{2m}\left[\left(\gamma_1 + \frac{5}{2}\gamma_2\right)k^2
	- 2 \gamma_2 \left(k_{x^\prime}^2 J_{x^\prime}^2 + k_{y^\prime}^2 J_{y^\prime}^2 + k_{z^\prime}^2 J_{z^\prime}^2\right)
	- \gamma_3 \left( \{k_{x^\prime}, k_{y^\prime}\} \{ J_{x^\prime}, J_{y^\prime}\} + \textrm{c.p.} \right)
	\right]\,
\end{equation}

where $\gamma_{1,2,3}$ are the Luttinger parameters, $m$ is the electron mass, $J_i$ are the spin-$\frac{3}{2}$ matrices, $k^2=k_{x^\prime}^2 + k_{y^\prime}^2 + k_{z^\prime}^2\,$, ``c.p.'' stands for cyclic permutation, and $\{A, B\}=AB + BA$ is the anticommutator.

For Ge, \mbox{$(\gamma_3 - \gamma_2) / \gamma_1 = 10.8\%$} and one can use the so called spherical approximation \cite{Kloeffel_DirectRashbaspinorbit_2017}.
By setting $\gamma_2 = \gamma_3 = \gamma_s$, where $\gamma_s = (2\gamma_2 + 3\gamma_3) / 5$, in Eq~\eqref{eq-supp-lk} the Hamiltonian becomes spherically symmetric,
and we may use a simulation coordinate system with $z$ being the direction along the nanowire and the cross section laying in the $xy$-plane.

\paragraph{Magnetic field}
We include magnetic field through the Zeeman term~\cite{Luttinger_QuantumTheoryCyclotron_1956, Winkler_SpinOrbitCouplingEffects_2003}
\begin{equation}
	H_{\textrm{Z}} = 2 \kappa \mu_B \boldsymbol{B} \cdot \boldsymbol{J}
\end{equation}
and the orbital effect by the following substitution in the Hamiltonian
\begin{equation}
	\boldsymbol{k} \rightarrow \boldsymbol{k} + \frac{2\pi}{\phi_0}\boldsymbol{A}\,,
\end{equation}
where $\boldsymbol{k} = (k_x, k_y, k_z) = -i\nabla$, $\boldsymbol{A}$ the vector potential, $\phi_0=\frac{h}{e}$ is the flux quantum with $e$ being positive elementary charge and $h$ the Planck constant.
The anisotropic Zeeman term $2q \mu_B \boldsymbol{B}\cdot\boldsymbol{\mathcal{J}}$~\cite{Winkler_SpinOrbitCouplingEffects_2003, Luttinger_QuantumTheoryCyclotron_1956}, where $\boldsymbol{\mathcal{J}}_i = J_i^3$, is omitted as $|q|<<|\kappa|$ for Si and Ge~\cite{Kloeffel_DirectRashbaspinorbit_2017, Lawaetz_ValenceBandParametersCubic_1971}.

\paragraph{Electric field} We include the electric field by the direct coupling~\cite{Kloeffel_DirectRashbaspinorbit_2017, Kloeffel_Strongspinorbitinteraction_2011} to the electrostatic potential
\begin{equation}
	H_{\textrm{dir}} = -e \boldsymbol{E}\cdot \boldsymbol{r}\,,
\end{equation}
We also consider indirect coupling originating from higher bands, excluded from the LK Hamiltonian, in form of a standard Rashba SOI term~\cite{Winkler_SpinOrbitCouplingEffects_2003}
\begin{equation}
	H_{\textrm{R}} = \alpha \boldsymbol{E} \cdot \boldsymbol{k} \times \boldsymbol{J}\,,
\end{equation}
where $\boldsymbol{E}$ is electric field and $\alpha$ is the Rashba coefficient.

\paragraph{Strain effect}
In our numerical calculations we only simulate the Ge core, and include the presence of the Si shell through the strain that it induces in the core \cite{Kloeffel_Strongspinorbitinteraction_2011}.
We model the strain using the Bir-Pikus Hamiltonian~\cite{Bir_Symmetrystraininducedeffects_1974}
\begin{align}
	\label{eq-supp-bir-pikus}
	H_\textrm{BP} =& - \left(a+\frac{5b}{4}\right)\Big(\epsilon_{x^\prime x^\prime} + \epsilon_{y^\prime y^\prime} + \epsilon_{z^\prime z^\prime}\Big) \nonumber\\
	&+ b\Big(\epsilon_{x^\prime x^\prime} J_{x^\prime}^2 + \epsilon_{y^\prime y^\prime} J_{y^\prime}^2 + \epsilon_{z^\prime z^\prime} J_{z^\prime}^2\Big) \nonumber\\
	&+ \frac{d}{\sqrt{3}} \Big(\epsilon_{x^\prime y^\prime} \{J_{x^\prime}, J_{y^\prime}\} + \textrm{c.p.}\Big)\,,
\end{align}
where $a\,, b\,, c$ are the deformation potentials and $\epsilon_{ij} = \epsilon_{ji}$ are the strain tensor elements.
Similarly to the Luttinger-Kohn Hamiltonian the spherical approximation can be used and strain may assumed to be constant in the Ge core~\cite{Kloeffel_Acousticphononsstrain_2014}. Thus, $d=\sqrt{3}b$, $\epsilon_\perp = \epsilon_{xx} = \epsilon_{yy}\,$, and $\epsilon_{xy} = \epsilon_{xz} = \epsilon_{yz}=0\,$. 
The Bir-Pikus Hamiltonian then simplifies to the spherical symmetric form~\cite{Kloeffel_Acousticphononsstrain_2014}

\begin{equation}
	\label{eq-supp-BP-invariant}
	H_\textrm{BP} = b (\epsilon_{zz} - \epsilon_\perp) J_z^2\,,
\end{equation}
where a global energy shift has been omitted.

\subsubsection{Material parameters}

Material parameters used in the simulation are given in Table~\ref{table-supp: Ge}.
We take the structure parameters for Ge from Ref.~\citenum{Lawaetz_ValenceBandParametersCubic_1971}, the effective Rashba coefficient from Ref.~\citenum{Kloeffel_DirectRashbaspinorbit_2017}, the deformation potentials from Refs.~\citenum{Bir_Symmetrystraininducedeffects_1974} and ~\citenum{Kloeffel_Acousticphononsstrain_2014}, and the strain parameters of the sample from Ref.~\citenum{Conesa_2017_SM}.
\begin{table*}[!htb]
	\caption{\label{table-supp: Ge}
	Band structure parameters for Ge~\cite{Kloeffel_DirectRashbaspinorbit_2017,Lawaetz_ValenceBandParametersCubic_1971} and strain parameters ~\cite{Bir_Symmetrystraininducedeffects_1974,Conesa_2017_SM}.
	All parameters are for $T=0$ K.}

\begin{tabular}{c | c | c | c | c | c | c | c | c | c }
	$\gamma_1$ & $\gamma_2$ & $\gamma_3$ & $\gamma_s$ & $\kappa$ & $\alpha$ [$\textrm{nm}^2e$] & b [eV] & d [eV]& $\epsilon_{zz}$ & $\epsilon_{rr}$  \\\hline
	$13.25$ & $4.25$ & $5.69$ & $5.114$ & $3.13$ & $-0.4$ & $-2.5$ & $-5.0$ & $-1.5$ & $3.5$
\end{tabular}
\end{table*}

\subsubsection{Numerical method}
We perform our numerical calculations using Kwant~\cite{Groth_Kwantsoftwarepackage_2014}.
We use the finite difference method to discretize the Hamiltonian~\eqref{eq-supp-hamiltonian} on a cubic grid with spacing $a\,$.
Depending on the geometry we use two slightly different methods.

The first approach is suitable for simulating a translation invariant infinite wire system, by considering the Hamiltonian
\begin{equation}
	H(k_x = -\textrm{i}\partial_x, k_y = -\textrm{i}\partial_y, k_z).
\end{equation}
The transverse momenta $k_x$ and $k_y$ are treated as differential operators, which are discretized as finite difference operators.
The Hamiltonian is then represented in a tight-binding form, and a finite system in the $xy$-plane is generated that represents the wire cross section.
The cross section has either a square or hexagon shape.
The momentum along the wire, $k_z\,$, remains a scalar parameter.

In the second approach we treat all momenta as differential operators:
\begin{equation}
	H(k_x = -\textrm{i}\partial_x, k_y = -\textrm{i}\partial_y, k_z = -\textrm{i}\partial_z).
\end{equation}
In addition to a finite cross section in the $xy$-plane we terminate the wire in the $z$-direction, effectively obtaining a quantum dot of length $L$.

The Land\'{e} $g$-factors are extracted from the energy spectrum of the system as a split in energies caused by the finite magnetic field
\begin{equation}
	\Delta E_n = g_n \mu_B B\,,
\end{equation}
where $n$ is band number, $\mu_B$ is Bohr magneton, and $B$ is the magnitude of the magnetic field.
For the infinite wire we use the energy split at $k_z=0\,$.
We note that this numerical approach goes beyond the effective Hamiltonian approach in Ref.~\citenum{Kloeffel_Strongspinorbitinteraction_2011} and also
takes into account the effects of higher states. The accuracy of our approach is controlled by the grid spacing $a$.

\subsubsection{Model geometry and verification}

The nanowires used in our experiment have a hexagonal cross-section with a corner-to-corner width of $6\,$nm. Faithfully representing this shape with a cubic lattice requires a rather small lattice spacing $a$ that is computationally unfavorable.
\begin{figure}[!ht]
	\center
	\includegraphics[width=.9\columnwidth]{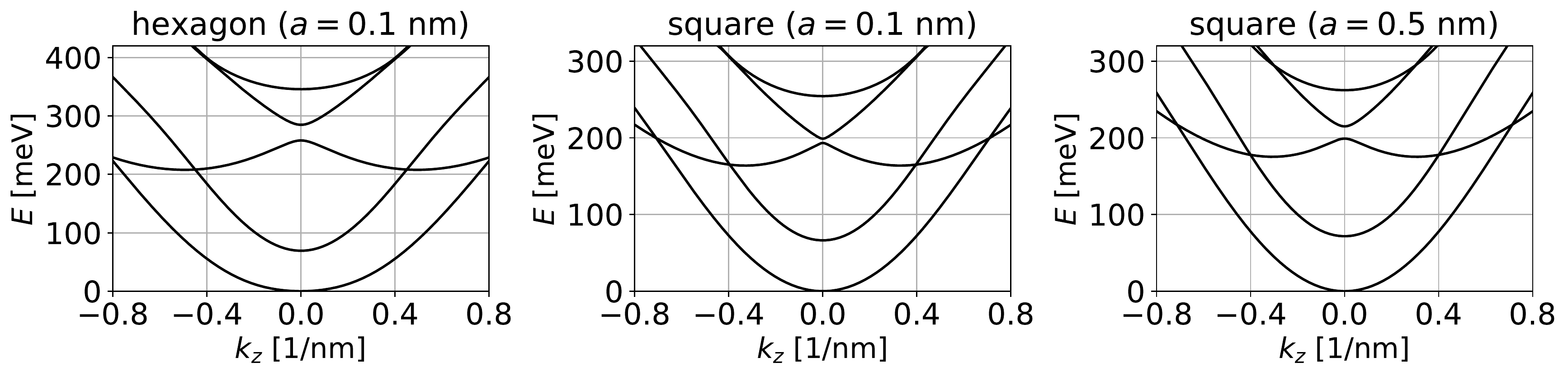}
	\caption{
	Dispersion of an infinite wire with hexagon (left) and square (middle, right) cross sections for different discretization grid spacings $a=0.1$ nm (left, middle) and $a=0.5$ nm (right).
	\label{fig:sup_infinite_wire_1a}}
\end{figure}
Figure~\ref{fig:sup_infinite_wire_1a} shows the comparison of the band structure between the wires with hexagon (left) and square (middle) shaped cross sections calculated using the grid spacing $a=0.1$ nm. We observe that the impact of the cross section shape on the qualitative result is small, in agreement with what was reported for the comparison between a circular and square cross section\cite{Kloeffel_DirectRashbaspinorbit_2017}. Hence we use a square cross section with $6\,$nm side length in further calculations.

This choice allows us to use a larger grid spacing ($a=0.5$ nm) that significantly reduces the computational cost of the calculation. For grid spacing $a=0.5$ nm the square cross section preserves the symmetries of the system and key features of the dispersion of two lowest subbands, see middle and right panel on Fig.~\ref{fig:sup_infinite_wire_1a}. We also note that the band structures we observe agree qualitatively with what was reported earlier \cite{Kloeffel_Strongspinorbitinteraction_2011, Kloeffel_DirectRashbaspinorbit_2017},
further verifying the accuracy of our approach.

\subsection{Simulation code and dataset}
All simulation codes used in this project are available under (simplified) BSD licence together with raw simulation data~\cite{data}.

\subsection{Results}

\subsubsection{Infinite wire}
We first investigate the infinite wire system.
In Fig.\,S5 we present the anisotropy of the $g$-factors when a magnetic field is included only through the Zeeman term, only the orbital effect, and both of these contributions, respectively.
The direction of magnetic field changes from along the $+z$ axis (parallel to the wire) to the $-z$ axis (antiparallel to the wire).
No electric field is present in the system.
The results show that the $k_z = 0$ states behave differently in the lowest two subbands.
In the lowest state the anisotropy originates almost exclusively from the Zeeman term.
On the other hand, in the second state the Zeeman and orbital contributions both have significant anisotropies but opposite signs, such that they partially cancel (note that the graphs show absolute values of the $g$-factors).
Comparing to Ref.~\citenum{Kloeffel_Strongspinorbitinteraction_2011} we find increased g-factor values, such as an order-of-magnitude enhancement of $g_z$ in the lowest suband. We can attribute this to strain, as our numerical simulations yield g-factor values comparable to Ref.~\citenum{Kloeffel_Strongspinorbitinteraction_2011} in the absence of strain. Also, strain leads to $g_z > g_{x,y}$ in the second subband, reversing the anisotropy. Note also that our results for the lowest subband agree better with the results of Ref.~\citenum{Kloeffel_DirectRashbaspinorbit_2017} with a somewhat weaker strain than in our situation.

\begin{figure}[!ht]
	\center
	\includegraphics[width=.9\columnwidth]{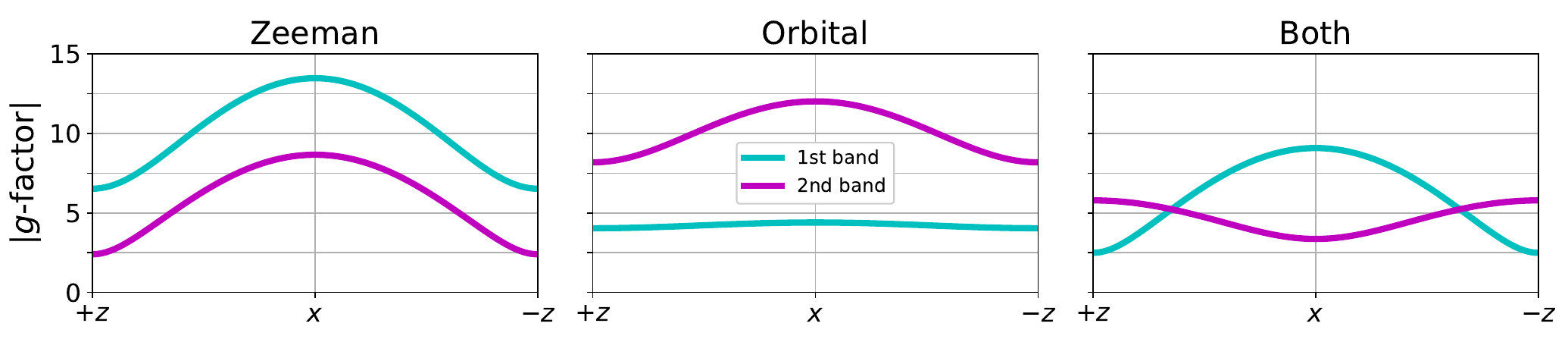}
	\caption{
	Anisotropy of the $g$-factors of the two lowest states in an infinite wire with square cross section.
	The magnetic field is included through the Zeeman term only, the orbital contribution only, and through both on the left, middle, and right panel respectively.
	The direction of the magnetic field changes from parallel to antiparallel to the wire.
	No electric field is present in the system.
	\label{fig:sup_infinite_wire_1b}}
\end{figure}

In Fig.\,S6 we analyse the anisotropy as the magnetic field orientation changes with respect to the wire from the parallel to antiparallel direction (left and middle) and around the perpendicular directions (right), in the absence and presence of electric fields.
The magnetic field is rotated from $+z$ to $-z$ axis through $+x$ (left) and $+y$ (middle).
In the right panel magnetic field changes from $+y$ through $+x$ to $-y$.
The upper row corresponds to systems with no electric field whereas the bottom row corresponds to systems with perpendicular electric field $E_x=10 \textrm{V}/\mu\textrm{m}\,$ that
we estimate for our experimental situation. Due to the fourfold rotational symmetry, $g$-factors are identical for $x$ and $y$ directions in the abscence of electric field as expected. This symmetry is in principle broken by the applied field, but the anisotropy between $x$ and $y$ remains small for experimentally relevant field strengths.

\begin{figure}[!ht]
	\includegraphics[width=.8\columnwidth]{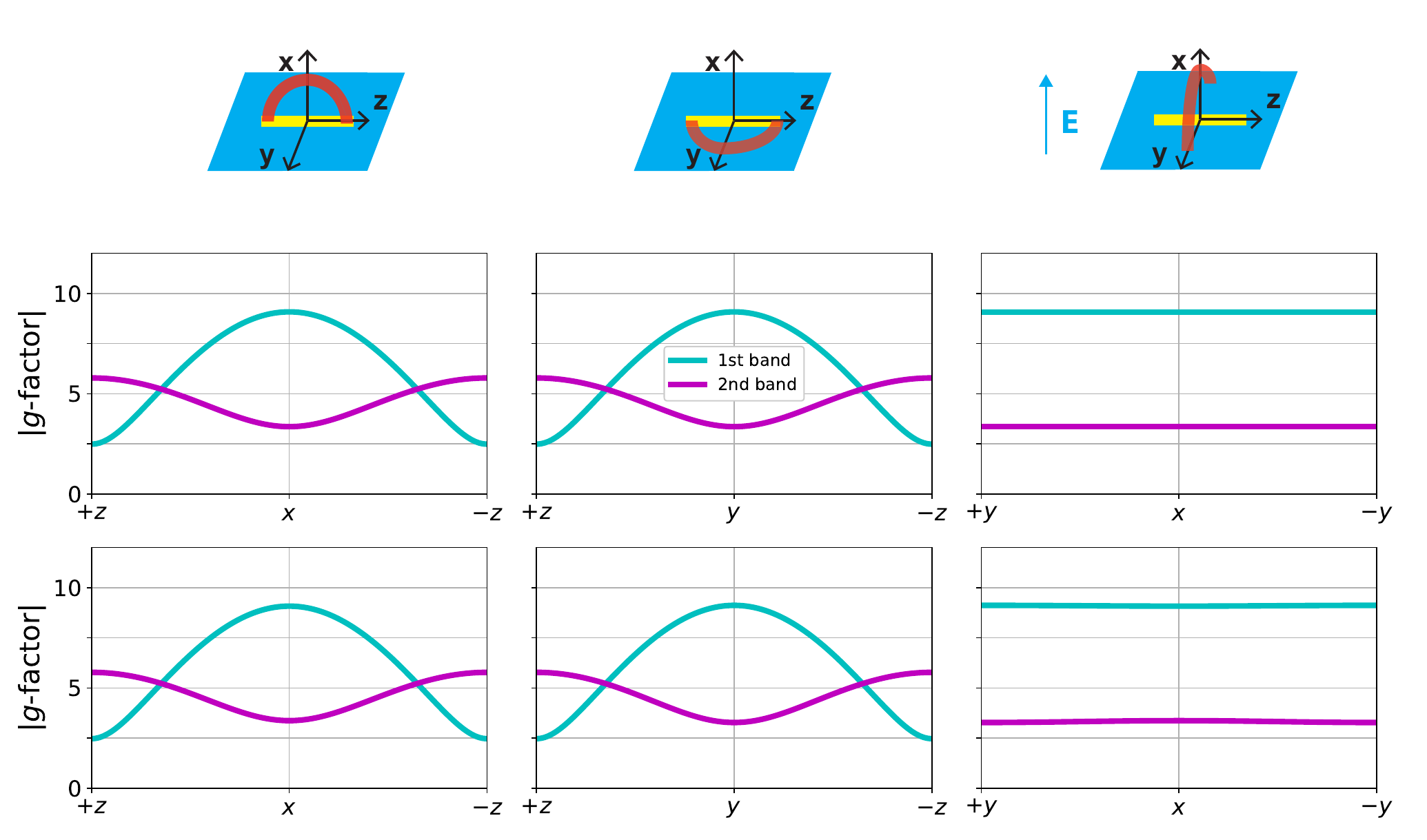}
	\caption{
	Anisotropy of the $g$-factors of the two lowest states in an infinite wire with square cross section.
	The upper panels correspond to zero electric field whereas on the lower panel a perpendicular electric field of $10\textrm{V}/\mu\textrm{m}$ is applied.
	\label{fig:sup_infinite_wire_2}}
\end{figure}

\begin{figure}[!h]
	\includegraphics[width=0.5\columnwidth]{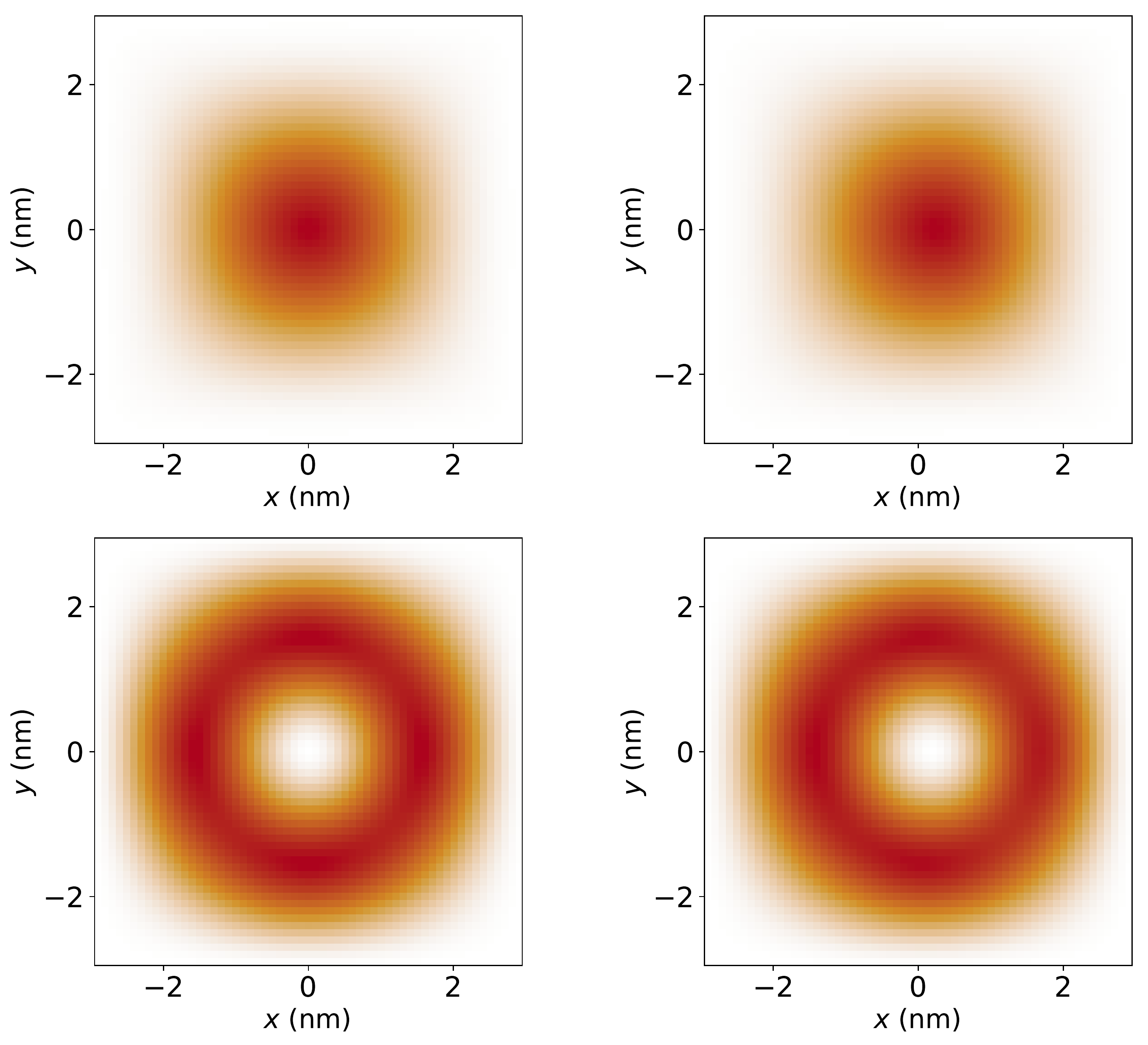}
	\caption{ The wavefunction of the two lowest bands for $k_z=0$. On the left panel we show the calculations done without electric field whereas on the right panel we show calculation under the electric field $10\textrm{V}/\mu\textrm{m}\,$.
	\label{fig:sup_wave-functions}}
\end{figure}

Indeed, we observe that due to the large confinement energy (around 80meV, see Fig.~\ref{fig:sup_infinite_wire_1a}) the effect of the electric field on the $k_z=0$ states in the infinite wire is almost negligible, as demonstrated in Fig.~\ref{fig:sup_wave-functions} (note that the shape of these wave functions in the absence of electric fields is in agreement with previous results \cite{Csontos_2007, Kloeffel_Strongspinorbitinteraction_2011, Kloeffel_DirectRashbaspinorbit_2017}).

In summary, we find that the g-factor anisotropy of the lowest subbands is modified considerably by strain. However, the results also do not agree with our experimental finding of a quenched anisotropy at lower densities. For this reason, we now turn to quantum dots.

\subsection{Quantum dot}

As explained in the main text, the experiment accesses higher states of the quantum dot, which originate from different subbands.
In this section we show results for a quantum dot of length $L=170$ nm (with hard-wall boundary conditions), corresponding to the experimental setup.
The discretization grid has $a = 0.5$ nm.

\begin{figure}[H]
	\includegraphics[width=.4\columnwidth]{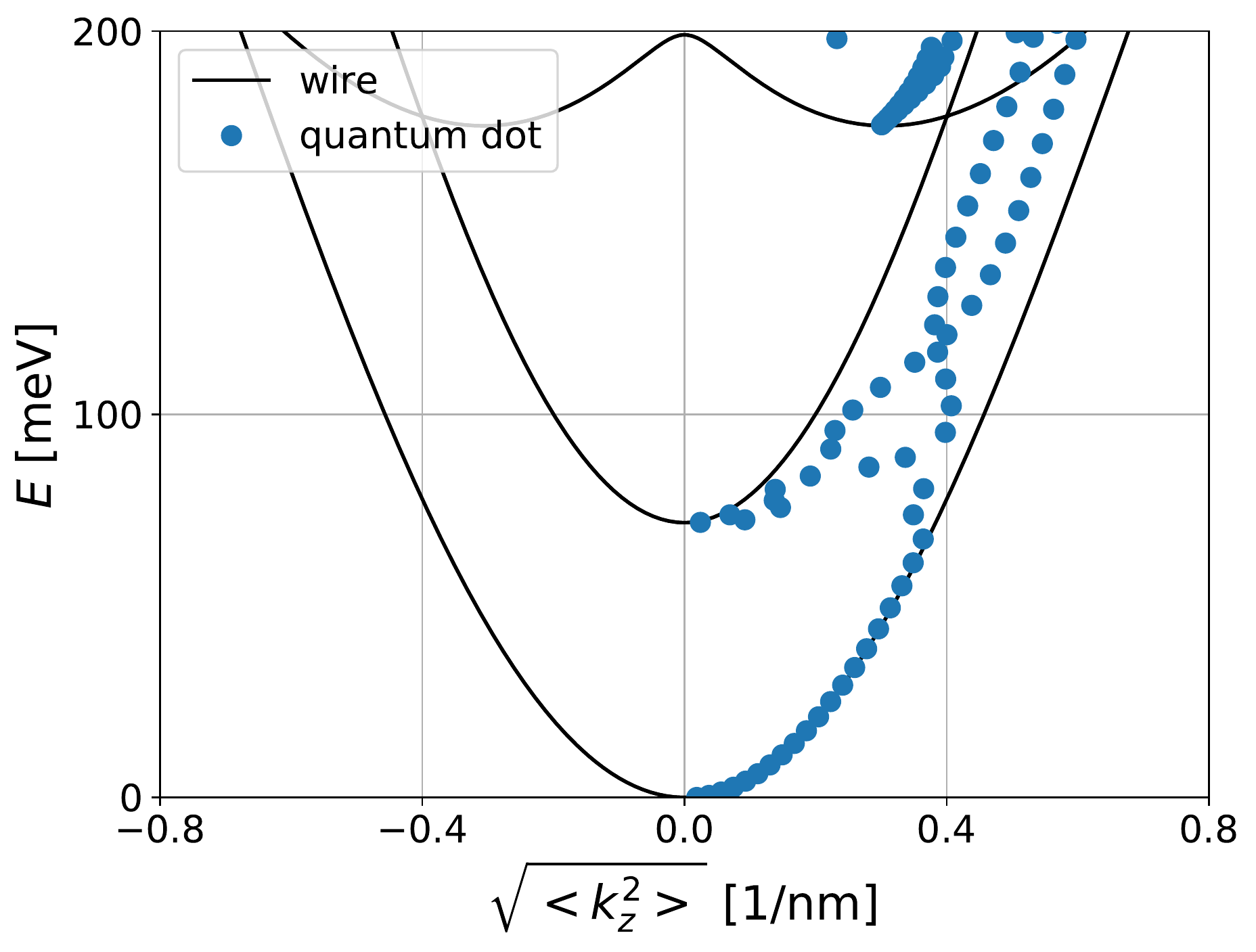}
	\caption{
	Quantum dot energies in function of $\sqrt{\langle k_z^2 \rangle}\,$ overlayed with the dispersion of the infinite wire.
	}
	\label{fig:sup_quantum_dot_dispersion}
\end{figure}

Fig.\,S8 shows the energy levels in the quantum dot as a function of $\sqrt{\langle k_z^2 \rangle}\,$ evaluated in the given eigenstate.
The states near the bottom of the lowest subband trace the infinite wire's dispersion very accurately, confirming the particle-in-box momentum quantization picture.
When the second subband enters, the quantum dot levels significantly deviate from the infinite wire dispersion.
This is a finite size effect, the result of mixing between states from different subbands with different $\langle k_z^2 \rangle\,$ (note that in a finite wire $k_z$ is not a conserved quantity).
For most of the energy window with two subbands the two branches of the dispersion are clearly distinguishable, supporting the view that consecutive quantum dot states inherit properties from different subbands.

\begin{figure}[H]
	\includegraphics[width=.8\columnwidth]{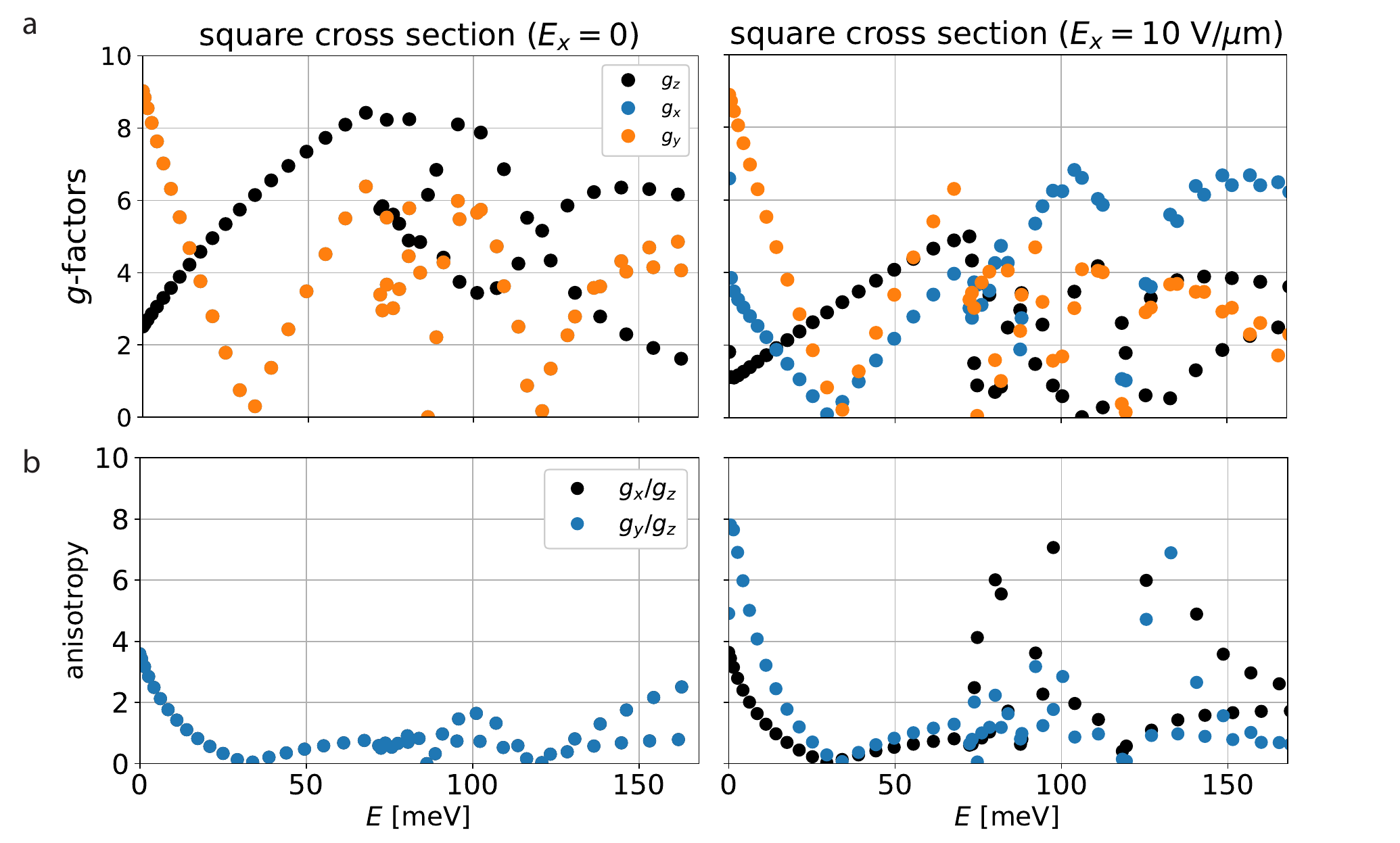}
	\caption{
	(a) $g$-Factors in the quantum dot of length $L=170$ nm with a square cross section. Without applied electric field $g_x$ and $g_y$ are identical.
	(b) $g$-Factor anisotropy calculated from the values found in (a).
	\label{fig:sup_quantum_dot}}
\end{figure}

Fig.\,S9a and Fig.\,S9b show the $g$-factors and $g$-factor anisotropies in the finite quantum dot respectively. 
At low energies, Fig.\, S9a reveals $g_{x,y}>g_z$ in the absence of an electric field ($E_x$) and $g_y >g_x$ in the presence of $E_x$. 
This is in qualitative agreement with previous calculations for the ground states in Ge-Si NW quantum dots~\cite{Maier_2013_SM}. Also, $g_y>g_x$ at finite $E_x$ has recently been observed experimentally~\cite{Brauns_2016}.
Where the second subband enters, the $g$-factor values split into two branches corresponding to the first and second subbands, this is especially visible in the $g_z$ values.
The external electric field induces a much larger anisotropy between $g_x$ and $g_y$ in higher states compared to the lowest one accessed at $k_z=0$, that was discussed previously in Ref.~\citenum{Maier_2013_SM}.  
Since the effect of the electric field on the g-factor in the infinite wire case is small, we attribute the increased anisotropy to spin-momentum locking present at nonzero $k_z$.

\pagebreak
\newpage

\providecommand{\latin}[1]{#1}
\makeatletter
\providecommand{\doi}
  {\begingroup\let\do\@makeother\dospecials
  \catcode`\{=1 \catcode`\}=2 \doi@aux}
\providecommand{\doi@aux}[1]{\endgroup\texttt{#1}}
\makeatother
\providecommand*\mcitethebibliography{\thebibliography}
\csname @ifundefined\endcsname{endmcitethebibliography}
  {\let\endmcitethebibliography\endthebibliography}{}

\end{document}